\documentclass[fleqn,usenatbib,useAMS]{mnras}

\usepackage{newtxtext,newtxmath}

\usepackage[T1]{fontenc}
\usepackage{ae,aecompl}

\usepackage{graphicx}	
\usepackage{amssymb}	
\usepackage{multicol}        
\usepackage{bm}		
\usepackage{enumitem}

\title[Radio emission from CIRs]{Radio Variability from Co-Rotating Interaction
Regions Threading Wolf-Rayet Winds}

\author[Ignace et al.]{Richard Ignace$^{1}$\thanks{E-mail: ignace@mail.etsu.edu}, Nicole St-Louis,$^{2}$ \&\ Raman K. Prinja,$^{3}$ \\\\
$^{1}$ East Tennessee State University, Department of Physics and Astronomy, Johnson City, TN 37614, USA\\
$^{2}$ D\'epartement de physique, Universit\'e de Montr\'eal, CP 6128, Succursale Centre-Ville, Montr\'eal, Qu\'ebec, H3C 3J7\\
$^{3}$ Department of Physics \&\ Astronomy, University College London, Gower Street, London WC1E 6BT, UK  \\
}

\date{Accepted 2020 July 6. Received 2020 July 6; in original form 2020 May 19}

\pubyear{2019}

\begin{document}
\label{firstpage}
\pagerange{\pageref{firstpage}--\pageref{lastpage}}
\maketitle

\begin{abstract}

The structured winds of single massive stars can be classified
into two broad groups:  stochastic structure and organized
structure.  While the former is typically identified with
clumping, the latter is typically associated with
rotational modulations, particularly the paradigm of
Co-rotating Interaction Regions (CIRs).  While CIRs have
been explored extensively in the UV band, and moderately in
the X-ray and optical, here we evaluate radio variability from
CIR structures assuming free-free opacity in a dense wind.  
Our goal is to conduct a broad parameter
study to assess the observational feasibility, and to this end,
we adopt a phenomenological model for a CIR that threads an
otherwise spherical wind.  We find that under reasonable assumptions,
it is possible to obtain radio variability at the 10\%
level.  The detailed structure of the folded light curve
depends not only on the curvature of the CIR, the density
contrast of the CIR relative to the wind, and viewing inclination,
but also on wavelength.  Comparing light curves at different
wavelengths, we find that the amplitude can change, that there can be phase shifts in the 
waveform, and the the entire waveform itself can change.  These characterstics could be exploited to detect the presence of CIRs in dense, hot winds.
\end{abstract}




\section{Introduction}
\label{Radio CIR intro}
Massive star ($M > 20 M_\odot$) winds input significant amounts of chemically enriched gas and mechanical energy into their surrounding interstellar medium (ISM), thereby greatly affecting the evolution of their host clusters and galaxies. Mass loss by winds also determines the ultimate fate of massive stars, and the nature of their (neutron star and black hole) remnants. Consequently, reliable measurements of mass-loss rates due to stellar winds are essential for all  these fields of study.

It is well-known that massive-star winds are driven by radiation pressure on metal lines \citep{1975ApJ...195..157C} but in recent years, it has become apparent that these winds are far more complex than the simple homogeneous and spherically symmetric flows originally envisioned. Instead, they have been shown to contain optically thick structures which may be quite small (micro-structures) or very large (macro-structures). Our understanding of stellar winds is at an important cross-road.  There is growing understanding that stellar rotation and magnetism play important but not yet understood roles in shaping the wind flow and generating emission across optical, UV, X-ray and radio wavebands. Until we unravel the details of these flows we cannot hope to accurately translate observational diagnostics into reliable physical quantities such as mass-loss rates. To progress, a firm grasp of the underlying physical mechanisms that determine the wind structures is needed. The state of affairs can be seen in recent literature in which the values of observationaly derived mass-loss rates have swung back and forth by factors of ten or more \citep{2006A&A...454..625P,2003ApJ...586..996M,2006ApJ...637.1025F,2011A&A...528A..64S,2012A&A...541A..37S}.


The current radiation-driven wind models show that for small-scale
wind structures, the line de-shadowing instabilities
\citep{1980ApJ...241.1131C,1979ApJ...231..514M,1984ApJ...284..337O} is
thought to play an important role in the development of small-scale
clumping, now known to be an ubiquitous attribute of hot-star winds
\citep{2008cihw.conf...17M}. For larger, spatially coherent structures,
\cite{1984ApJ...283..303M,1986A&A...165..157M} suggested that spiral
shaped Co-rotating Interaction Regions (CIRs) could be relevant.
In 2D hydrodynamic simulations, \citet{1996ApJ...462..469C} showed
that when perturbations in the form of bright or dark spots are
present on the surface of a massive star, corotating structures
develop when flows from a rotating star accelerating at different
rates collide.

The CIR model was successful in reproducing IUE ultraviolet (UV)
spectroscopic timeseries of $\xi$ Per O7.5 III(n)((f))
\citep{2001A&A...368..601D}, as well as HD64760, B0.5 Ib
\citep{1997A&A...327..699F}, and its signature appears to be present
in most UV spectroscopic timeseries available for O stars.  It
predicts spiral structures consisting of density enhancements of
$\sim$2 for a radiative force enhancement of 50\% (e.g., owing to
a bright spot), together with velocity plateaus which can increase
the Sobolev optical depth by factors of 10 to 100. Detailed 3D
radiative transfer and hydrodynamic calculations by
\cite{2008ApJ...678..408L} for HD64760 led to density contrast
increases for the CIR of 20\%-30\%\ and opening angles of
20$^\circ$-30$^\circ$. The passage of CIR spiral arms across the
line-of-sight to the stellar disk leading to Discrete Absorption
Components (DACs) and/or the formation of propagating discontinuities
in the velocity gradient forming Periodic Absorption Modulations
(PAMs) account for the wind UV P Cygni absorption component
variability, and modeling indicates that observed DACs are best explained
in terms of a paradigm involving bright spots to drive CIR
structures, as opposed to dark spots \citep[e.g.,][]{2017MNRAS.470.3672D}.

For very dense winds such as those of Wolf-Rayet (WR) stars, the
UV P Cygni absorption troughs are usually saturated, which prevents
the detection of such variability.  One exception is the WN7 stars,
HD93131 (WR24) for which \citet{1992A&A...266..377P} found a migrating
DAC in the He{\sc ii}$\lambda$1640 P Cygni profile.  Therefore,
evidence for CIRs must be searched for in emission lines instead.
\citet{2002A&A...395..209D} carried out theoretical calculations
for optically thin emission-line variability for a radiatively-driven
wind in the presence of CIRs.  They predict an unambiguous S-shape
variability pattern in dynamic spectra illustrating line-profile
variability as a function of time.  Such a variability pattern has
been found in several optical emission lines of a few WR stars such
as HD50896 \citep[WR6: e.g.][]{1997ApJ...482..470M}, HD191765
\citep[WR134: e.g.][]{2016MNRAS.460.3407A} and HD4001 \citep[WR1:
e.g.][]{2010ApJ...716..929C}.The CIRs can undoubtedly strongly
affect the observational diagnostics used to determine the true
mass-loss rates.

The CIR density enhancements also provide a potentially powerful,
but untested, means for producing radio variability.  Hot star winds
emit radio radiation through (thermal) free-free emission, due to
electron-ion interactions in their ionized wind
\citep{1975MNRAS.170...41W}. The density squared dependence of the
free-free flux makes the radio observations extremely sensitive to
clumping and density enhancements in the wind.

If a large-scale structure such as a CIR is present in the wind,
the projected area of the effective radio photosphere of the star
will be altered. Variability from a CIR will derive from a phase
dependence of the projected photosphere with stellar rotation.
Indeed, a CIR is like an asymmetric appendage and assuming it is a
high density region compared to the ambient wind, then the effective
projected radio photosphere will appear to have an extension to one
side. Essentially, if the CIR is denser than the ambient wind, it
will generate a sector of extended radio photosphere, relative to
a wind with no CIR.  Modulation of the radio photosphere with
rotational phase will lead to periodic continuum flux variations
for the unresolved source as long as the structure in unchanged.
Indeed, the consideration is much in the same spirit as applications
for resolved dusty spiral structures that form in massive star
colliding wind binaries
\citep[e.g.,][]{2007ApJ...655.1033M,2016MNRAS.460.3975H}.  The
differences are strong shocks, modulation on the orbital period of
the binary instead of rotational period of a star, and many dusty
spirals have been spatially resolved.

The paper is organized as follows.  Section~\ref{section2} provides a
review of the free-free opacity as used in calculating the radio
properties of spherical stars.  Based on this, the theory is
expanded to application for CIR structures.  Then Section~\ref{section3}
provides multi-wavelength lightcurves for a broad combination
of model parameters.  Concluding remarks and observational
prospects are presented in Section~\ref{section4}.


\section{Model Description}
\label{section2}

\subsection{Radio SED for a Spherical Wind}
\label{sub:sph}

The radio spectrum from thermal free-free opacity in
a dense and optically thick massive-star wind is
a well-known problem for spherical symmetry; in short, the effective radio
photosphere has an extent that grows with wavelength 
\citep{1975A&A....39....1P,1975MNRAS.170...41W}.
When sufficiently large, the emission swamps that of the star,
and the spectral shape is typically a power-law with 
wavelength, having a slope that is much more shallow
than Rayleigh-Jeans.

The inclusion of a CIR in such a framework significantly 
complicates the calculation.  There is generally a complete
loss of any geometrical symmetry, and the radiative transfer problem
must be handled numerically.  Plus the signal will
be rotationally modulated.  Our study of variable radio emission for a wind threaded by a CIR will
make use of several simplifying assumptions as our goal is to explore a fairly broad parameter space.

First among the simplifications is that we treat the wind
as optically thick such that a radio photosphere forms in the
outflow.  Direct emission from the star itself is assumed to be highly absorbed.  We further assume that the wind is isothermal and that the ionization
of atomic species is fixed.  Such assumptions could be relaxed,
with the effect of altering the shape of the SED and its luminosity.
None of these factors produce variability unless they are
themselves time-varying.  However, allowing for radius-dependence
in the temperature or ionization properties would introduce additional
free parameters into the calculation that are not germane to the question of
how CIRs influence radio emission and variability.  The
above simplifications are made for convenience not necessity,
since the goal is to explore how geometry can drive the variable
signal.

At this point it is important to identify an appropriate fiducial
against which to consider the variable radio flux.   To this end,
the natural comparison is the case of a strictly spherical wind
with no CIR.  While the solution for the thermal free-free emission
is well known, following \cite{2009AN....330..717I}
and \cite{2016MNRAS.457.4123I},
we briefly review the steps here to provide a backdrop
for modification when a CIR structure is included in \S~\ref{sub:cir}.

We begin with specifiying the free-free opacity in the Rayleigh-Jeans
limit of $h\nu \ll kT$ as given by \citep{2000asqu.book.....C}:

\begin{equation}
\kappa_\nu \rho = 0.018\, \frac{Z_{\rm i}^2}{\mu_{\rm i}\,\mu_{\rm e}}\,
        \frac{\rho^2}{m_H^2}\,T^{-3/2}\,g_\nu\,\nu^{-2} \, {\rm cm}^{-1},
\end{equation}

\noindent where $Z_{\rm i}$ is the rms ion charge, $\mu_{\rm i}$
and $\mu_{\rm e}$ are mean molecular weights per free ion and per
free electron, respectively, $\rho$ is the mass density of the gas,
$m_H$ is the mass of a hydrogen atom, $T$ is the gas temperature, $g_\nu$
is the free-free Gaunt factor, and $\nu$ is the frequency of
observation. All variables are in cgs units. 

The solution for the emission flux requires introduction of the
optical depth, $\tau_\nu$.  The optical depth from a distant observer
to a point in the wind that lies along the line-of-sight to
the star center is given by

\begin{equation}
\tau_\nu = \int_r^\infty \, \kappa_\nu \rho\,dr.
\end{equation}

\noindent Allowing for microclumping parameterized in terms of a
volume filling factor of density, $f_V$, the radial optical depth
of the preceding equation can be written as

\begin{equation}
\tau_\nu = \tau_0(\lambda)\,\int_{\tilde{r}}^\infty\,\frac{1}{f_V(\tilde{r})}\,
        \left[\frac{\rho(\tilde{r})}{\rho_0}\right]^2\,d\tilde{r},
        \label{eq:los_tau}
\end{equation}

\noindent with $\tilde{r}=r/R_\ast$, where $R_\ast$ is the stellar radius,
$\rho_0$ is the density at the base of the wind, and $\tau_0$
is a characteristic optical depth scale as a function of wavelength.
The latter is given by

\begin{equation}
\tau_0(\lambda)  =  5.4\times 10^{25}\, \frac{Z_{\rm i}^2}{\mu_{\rm i}\,
        \mu_{\rm e}}\,g_\nu\,
        T_{\rm kK}^{-3/2}\,   \left(\frac{\dot{M^2}}{R_\ast^3\,v_\infty^2}\right)\,\lambda_{\rm cm}^2,
\end{equation}

\noindent where $T_{\rm kK}$ is the wind temperature in kK, $\dot{M}$
is the mass-loss rate in $M_\odot$ yr$^{-1}$, $v_\infty$ is the wind
terminal speed in km s$^{-1}$, and $R_\ast$ is the stellar radius
in $R_\odot$.  Here the stellar radius refers to the wind base, or
where the gaseous layers transition from hydrostatic equilibrium
to the wind.

From \cite{1975MNRAS.170...41W} and \cite{1975A&A....39....1P}, 
key results for the emergent intensity and unresolved radio flux
are summarized in the following.  For an observer
sightline at impact parameter $\tilde{p} = p / R_\ast$ through the
spherical wind, the intensity is

\begin{equation}
I_\nu(\tilde{p}) = B_\nu(T)\,\left[1-e^{-\tau_{\rm tot}(\tilde{p})}\right],
\end{equation}

\noindent where wind is taken as isothermal, $\tau_{\rm tot}$ is
the total optical depth of the wind along a ray of impact
parameter $\tilde{p}$, and $B_\nu$ is the Planck function.  The total
optical depth is given by the integral

\begin{equation}
\tau_{\rm tot}(\tilde{p})= \tau_0(\lambda)\,\int_{-\infty}^{+\infty}\,\frac{1}{f_V(\tilde{r})}\,
        \left[\frac{\rho(\tilde{r})}{\rho_0}\right]^2\,d\tilde{z}.
\end{equation}

\noindent For a star of radius $R_\ast$ at a distance $D$ from
Earth, the expected flux of radiation from the wind becomes

\begin{equation}
F_\nu = 2\pi\,\frac{R_\ast^2}{D^2}\,B_\nu(T)\,\int_0^\infty\,
        \left[1-e^{-\tau_{\rm tot}(\tilde{p})}\right]\,\tilde{p}\,d\tilde{p}.
\end{equation}

Note that the preceding expression is for the wind emission only.
The total flux should account for emission by the stellar atmosphere,
which is attenuated by the wind opacity, and also stellar occultation
of wind emission.  However, our application is for the situation
when the radio photosphere is relatively large compared to the
stellar radius, and the hydrostatic atmosphere is strongly absorbed,
while the influence of occultation is small.

The case of a large radio photosphere also implies that the wind
is optically thick out to where the flow expands at the terminal
speed.  Adopting $v \approx v_\infty$, the wind density is an inverse
square law.  Under these conditions, both the optical depth
and radio flux become analytic.
The optical depth to any point in the spherical
wind becomes

\begin{equation}
\tau(\tilde{p},\theta) = \frac{\tau_0(\lambda)}{2\tilde{p}^3}\, \left[
        \theta - \frac{1}{2}\,\sin (2\theta) \right],
        \label{eq:tsmooth}
\end{equation}

\noindent where $\tilde{z} = \tilde{p}/\tan\theta$ and $\tilde{r}
= \tilde{p}/\sin\theta$.  Note that $\tau_{\rm tot}$ is achieved
when $\theta$ goes to $\pi$, which gives $\tau_{\rm tot} =
\pi\,\tau_0(\lambda)/2\tilde{p}^{3}$.

\citet{1977ApJ...212..488C} described how free-free flux that forms
in the wind could be interpreted in terms of a pseudo-photosphere.
Their argument was to evaluate an effective radius for the radio
photosphere, and it is useful to consider this scaling.
Here we adopt the notation that $r_1$ is where $\tau_\nu =
1$ along the line-of-sight.  From equation~(\ref{eq:los_tau}), one obtains

\begin{equation}
r_1 = \left[\frac{\tau_0(\lambda)}{3\,f_V}\right]^{1/3}\, R_\ast \propto
        g_\nu^{1/3}\,\lambda^{2/3}\,f_V^{-1/3}\, R_\ast.
\end{equation}

\noindent When $r_1 \gg R_\ast$, the flux integral becomes 

\begin{equation}
F_\nu \approx 2\pi\,\frac{R_\ast^2}{D^2}\,B_\nu(T_{\rm w})\,
        \int_0^\infty\,\left(1-e^{-\pi\,\tau_0/2\tilde{p}^3}\right)
        \,\tilde{p}\,d\tilde{p}.
\end{equation}

\noindent The analytic solution to this integral is

\begin{equation}
F_\nu = \Gamma\left(\frac{1}{3}\right)\times \,
        \frac{\pi\,R_\ast^2}{D^2}\,B_\nu(T_{\rm w})\,\left[
        \frac{\pi\tau_0(\lambda)}{2}\right]^{2/3},
        \label{eq:smoothflux}
\end{equation}

\noindent where $\Gamma$ is the ``Gamma'' function.  

With $g_\nu \propto \lambda^{0.11}$ in the radio band
\citep{2000asqu.book.....C}, the radio SED is a power law with a
logarithmic slope exponent of about $-0.6$.  Radio spectra observed
to display this power-law slope generally signals that the wind is
isothermal, spherical, and at terminal speed.  Slight deviations,
especially somewhat steeper negative slopes, may indicate variations
in the temperature or ionization of the wind.  Indeed for the
Rayleigh-Jeans limit, \cite{1977ApJ...212..488C} generalized their
results to relate an observed SED slope in terms of power-law
exponents for the density and temperature distributions.  If $\rho
\propto r^{-2}$, they showed that spherical winds have $F_\nu \propto
\lambda^{-0.6}$ even if the temperature varies as a power-law
distribution.

On the other hand, deviations from the standard power-law slope of
$-0.6$ can be an indicator of a variety of effects, such as ionization
gradients or that the free-free emission forms in the wind acceleration
zone, relevant for lower density winds.  A radio SED with positive
slope would be non-thermal, generally interpreted as related to
synchrotron emission and the presence of magnetism in the extended
wind \citep[e.g.,][]{1985ApJ...289..698W,2011BSRSL..80...67B}.

Application to winds threaded by CIRs must account for radio
variability. The preceding analysis for the radiative transfer must
be modified to take account of the non-spherical geometry, as
presented next.

\subsection{Radio Variability with a CIR}
\label{sub:cir}

\cite{1996ApJ...462..469C} and \cite{2017MNRAS.470.3672D} have
explored the structure of equatorial CIRs with 2D hydrodynamical
simulations.  \cite{2004A&A...423..693D} and \cite{2008ApJ...678..408L}
conducted 3D simulations for CIRs.  Both approaches employ the
concept of a bright starspot to establish a differential flow leading
to the spiral structure.  Here we employ a kinematic prescription
for a CIR to conduct a broad parameter study for radio variability.
For a CIR that emerges from the photosphere of the star with radius
$R_\ast$ corresponding to where the wind initiates, the CIR pattern
is provided by consideration of a ``streak line''
\citep[see][]{1996ApJ...462..469C}.  An expression for the geometric
center of the CIR is given by:

\begin{equation}
\varphi = \omega\, t + \varphi_0 - \frac{R_\ast\sin\,\vartheta_0}{r_0}\, \left[\frac{1-u}{u} + b \ln\left(\frac{w}{u\,w_0}\right)\right],
        \label{eq:kinematics}
\end{equation}

\noindent where $\varphi$ is the azimuth of the local center for
the CIR, $\varphi_0$ is a constant, $r$ is the radial distance in
the wind, $\vartheta_0$ is the co-latitude at which the CIR emerges
from the star, $u = R_\ast/r$ and $\omega$ is the angular speed of
rotation for the star (assumed solid body).  The factor $w$ is the
normalized wind velocity, $w=v(r)/v_\infty$, with $v_\infty$ the
wind terminal speed, and the wind velocity given by

\begin{equation}
v(r) = v_\infty\,\left(1-b\,u \right).
\end{equation}

\noindent for which the initial wind speed is $w_0 = 1-b$.
Finally, the parameter $r_0 = v_\infty/\omega$ is the
``winding radius'', which is a length scale associated with how
rapidly the CIR transitions to a spiral pattern.

\begin{figure}
\includegraphics[width=\columnwidth]{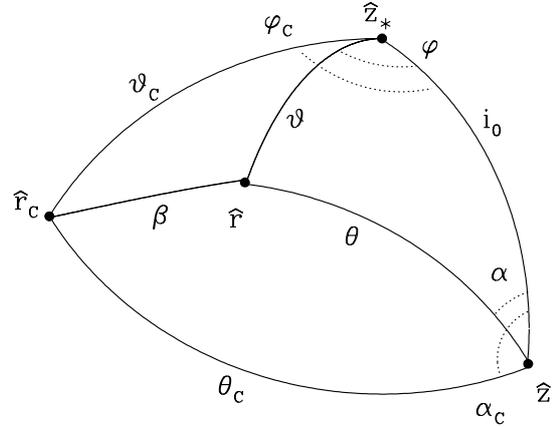}
 \caption{Geometry associated with calculating the emergent radio
 flux.  See text for definition of coordinates.}
  \label{F0}
\end{figure}

Figure~\ref{F0} shows the relevant geometry
associated with determining when a sightline
intersects the CIR.  The rotation axis of
the star is signified by the unit vector $\hat{z}_\ast$
with angular coordinates $\vartheta$ for co-latitude and
$\varphi$ for azimuth.
The observer is along $\hat{z}$, inclined by an angle
$i_0$.  The observer has angular coordinates of $\theta$ for
polar angle and $\alpha$ for azimuth.  For any radius
in the wind, a CIR will have a center point at that
distance, here indicated by $\hat{r}_C$.  The coordinates
for that point are $(\vartheta_C,\varphi_C)$.  As
noted, we will consider only a single equatorial CIR, 
for which
$\vartheta_C = 90^\circ$, and $\varphi_C(r)$ is the
solution given by equation~(\ref{eq:kinematics}).  Here
$\varphi_C(r)$ is the trace of the center of the spiral
feature.  In 
the observer frame, the location of this center
is given by $(\theta_C(r),\alpha_C(r))$.

The arc between $\hat{r}$ and $\hat{r}_C$ is $\beta$.  
The CIR has a half-opening angle $\beta_0$.  Whether
the point of interest falls within ($\beta\le \beta_0$)
or outside ($\beta>\beta_0$)
the CIR volume requires finding $\beta$, which
is given by

\begin{equation}
    \cos \beta = \cos\theta_C\cos\theta+
    \sin\theta_C\sin\theta\cos(\alpha_C-\alpha).
\end{equation}

\noindent With $\vartheta_C=90^\circ$, we have $\cos\theta_C=\sin i_0\cos \varphi_C(r)$.  The point along the ray is $(\theta,\alpha)$,
which are givens.  All that remains is determining $\alpha_C$.  
Using the law of cosines and the law of sines gives:

\begin{eqnarray}
0 & = & \cos \theta_C\cos i_0+\sin\theta_C \sin i_0\cos\alpha_C \\
\sin\theta_C\sin\alpha_C & = & - \sin\varphi_C .
\end{eqnarray}

\noindent Combining yields 

\begin{equation}
\tan \alpha_C = \frac{\tan\varphi_C}{\sin i_0}.  
\end{equation}

In summary, given a location $r,\theta,\alpha$ for observer
coordinates, the above coordinate transformations yield
both $\vartheta,\varphi$ and $\theta_C,\alpha_C$ to allow
evaluation of the angle $\beta$ between $\hat{r}$ and
$\hat{r}_C$.  Then $\beta$ can
be compared with the opening angle of the CIR to determine
whether a point, at a given time, falls within or outside
of the CIR structure. 

\begin{table}
\caption{Details of the various parameters used in our calculations of the various radio lightcurves for a single equatorial CIR in an optically thick wind.}
\begin{center}
\begin{tabular}{lccccc}
\hline\hline Figure & $\beta$ & $i$ & $\eta$ & $v_{\rm rot}$ \\
 & ($^\circ$) & ($^\circ$)  & & (km s$^{-1}$) \\ \hline
 &  &  &  &   \\
 \ref{F1}-left/top & 15 & 30 &  3 & 0 \\
 \ref{F1}-left/mid & 15 & 30 &  3 & 52 \\
 \ref{F1}-left/bot & 15 & 30 &  3 & 175 \\
 &  &  &  &   \\
 \ref{F1}-center/top & 15 & 60 &  3 & 0 \\
 \ref{F1}-center/mid & 15 & 60 &  3 & 52 \\
 \ref{F1}-center/bot & 15 & 60 &  3 & 175 \\
 &  &  &  &   \\ 
 \ref{F1}-right/top & 15 & 90 &  3 & 0 \\
 \ref{F1}-right/mid & 15 & 90 &  3 & 52 \\
 \ref{F1}-right/bot & 15 & 90 &  3 & 175 \\
 &  &  &  &    \\
 \ref{F2}-left/top & 15 & 30 &  9 & 0 \\
 \ref{F2}-left/mid & 15 & 30 &  9 & 52 \\
 \ref{F2}-left/bot & 15 & 30 &  9 & 175 \\
 &  &  &  &   \\
 \ref{F2}-center/top & 15 & 60 &  9 & 0 \\
 \ref{F2}-center/mid & 15 & 60 &  9 & 52 \\
 \ref{F2}-center/bot & 15 & 60 &  9 & 175 \\
 &  &  &  &    \\
 \ref{F2}-right/top & 15 & 90 &  9 & 0 \\
 \ref{F2}-right/mid & 15 & 90 &  9 & 52 \\
 \ref{F2}-right/bot & 15 & 90 &  9 & 175 \\
 &  &  &  &    \\
 \ref{F3}-left/top & 25 & 30 &  3 & 0 \\
 \ref{F3}-left/mid & 25 & 30 &  3 & 52 \\
 \ref{F3}-left/bot & 25 & 30 &  3 & 175 \\
 &  &  &  &    \\
 \ref{F3}-center/top & 25 & 60 &  3 & 0 \\
 \ref{F3}-center/mid & 25 & 60 &  3 & 52 \\
 \ref{F3}-center/bot & 25 & 60 &  3 & 175 \\
 &  &  &  &    \\
 \ref{F3}-right/top & 25 & 90 &  3 & 0 \\
 \ref{F3}-right/mid & 25 & 90 &  3 & 52 \\
 \ref{F3}-right/bot & 25 & 90 &  3 & 175 \\
 &  &  &  &    \\
 \ref{F4}-left/top & 25 & 30 &  9 & 0 \\
 \ref{F4}-left/mid & 25 & 30 &  9 & 52 \\
 \ref{F4}-left/bot & 25 & 30 &  9 & 175 \\
 &  &  &  &   \\
 \ref{F4}-center/top & 25 & 60 &  9 & 0 \\
 \ref{F4}-center/mid & 25 & 60 &  9 & 52 \\
 \ref{F4}-center/bot & 25 & 60 &  9 & 175 \\
 &  &  &  &    \\
 \ref{F4}-right/top & 25 & 90 &  9 & 0 \\
 \ref{F4}-right/mid & 25 & 90 &  9 & 52 \\
 \ref{F4}-right/bot & 25 & 90 &  9 & 175 \\ \hline
\label{tab1}
\end{tabular}
\end{center}
\end{table}

\begin{figure*}
\includegraphics[width=0.67\columnwidth]{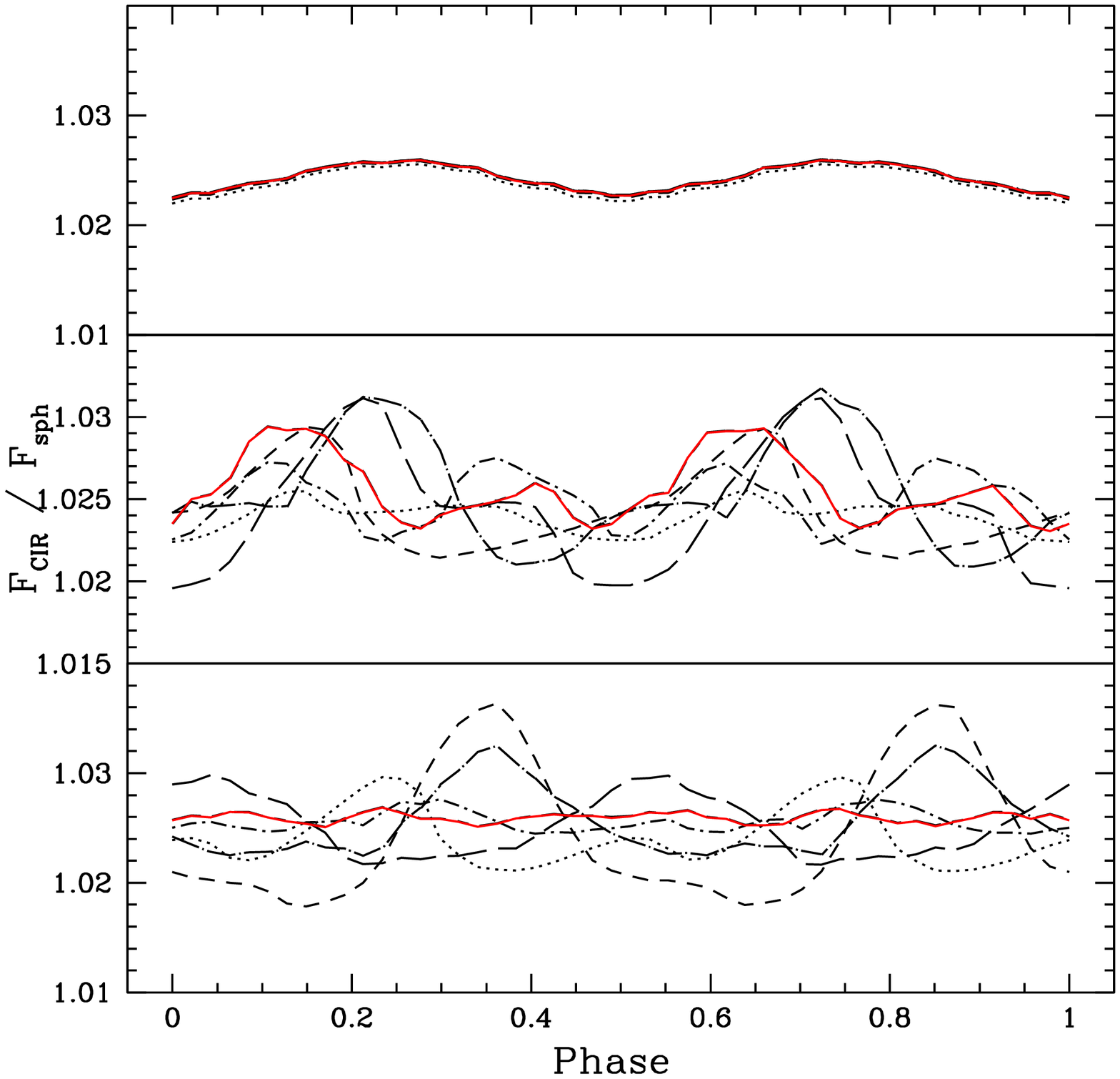}
\includegraphics[width=0.67\columnwidth]{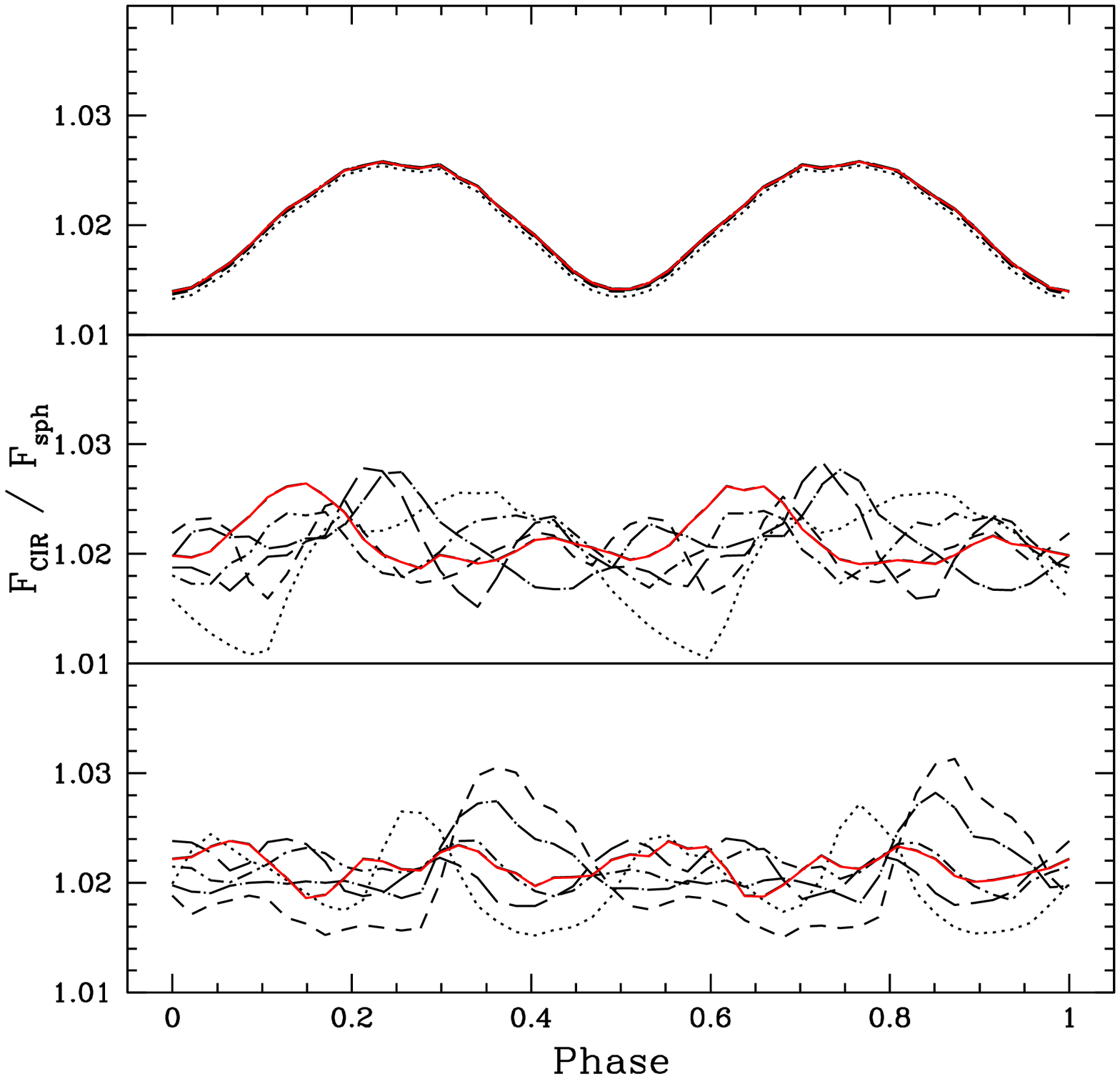}
\includegraphics[width=0.67\columnwidth]{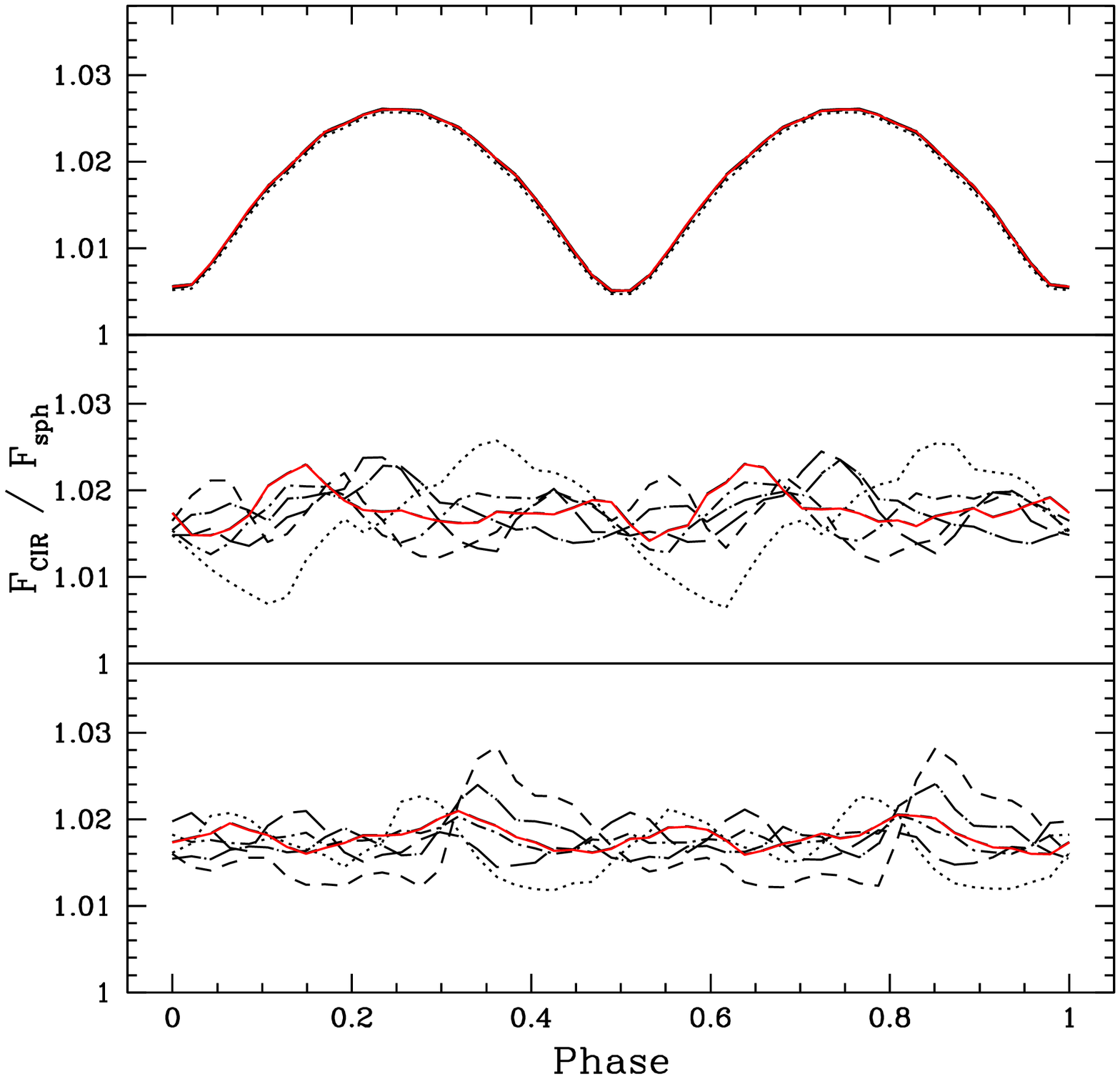}
 \caption{Model calculations for the radio flux as a function of rotational phase from a wind threaded by an
 equatorial CIR.  From left to right, the panels are for 3 different
 viewing inclinations of $i=30^\circ$, $60^\circ$, and $90^\circ$.  From
 top to bottom, 
the 3 panels are for different stellar rotation rates.  The
light curves are for different wavelengths, with red being the longest
 wavelength of the simulation at 31.6~cm and black dotted being the shortest at 1~mm (see text for more details).  The curves are fluxes normalized to
 that of a spherical wind at each respective wavelength and plotted with
 rotation phase.  Tab.~\ref{tab1} details specific parameters for each panel.}
  \label{F1}
\end{figure*}

\begin{figure*}
\includegraphics[width=0.67\columnwidth]{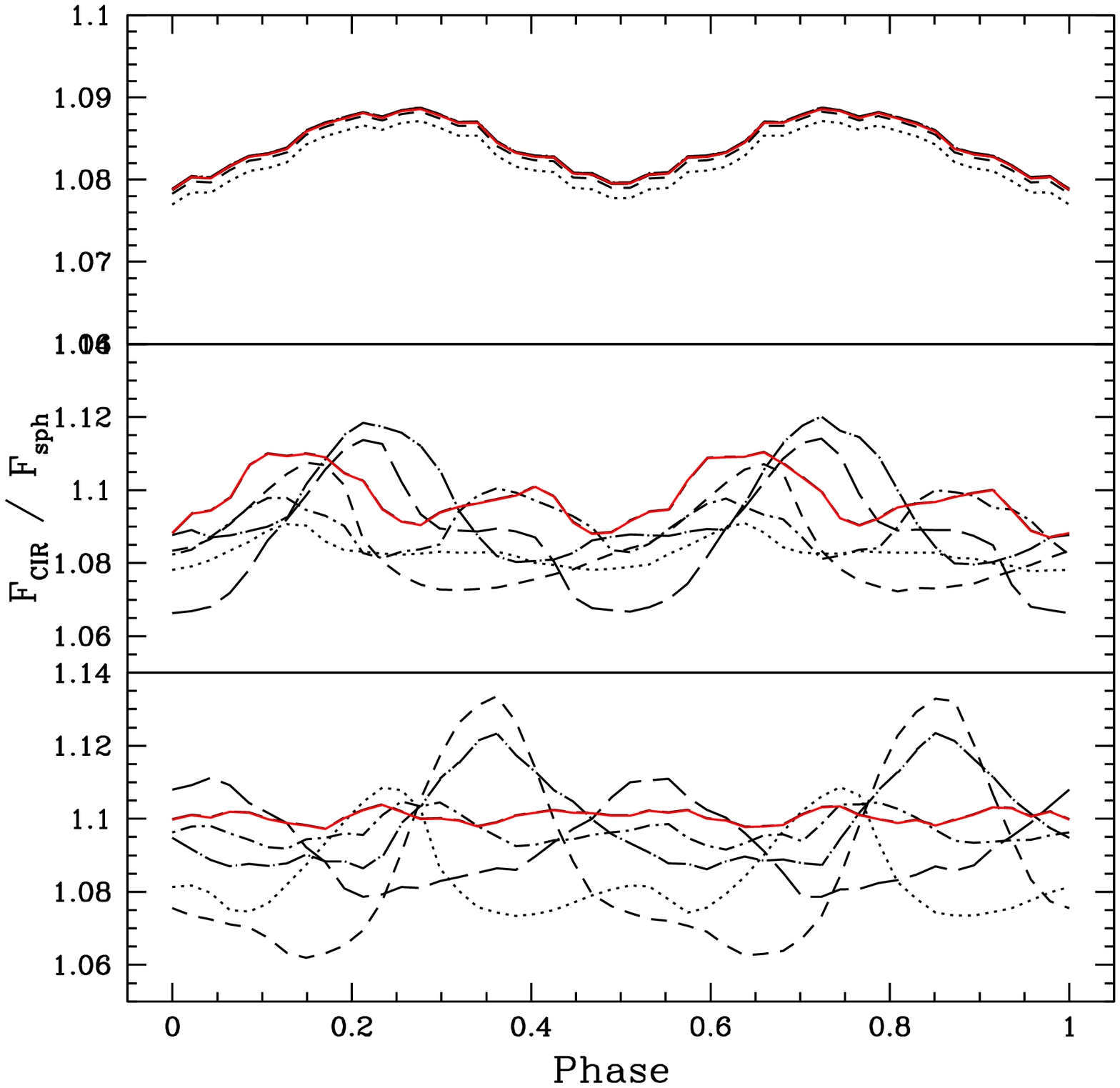}
\includegraphics[width=0.67\columnwidth]{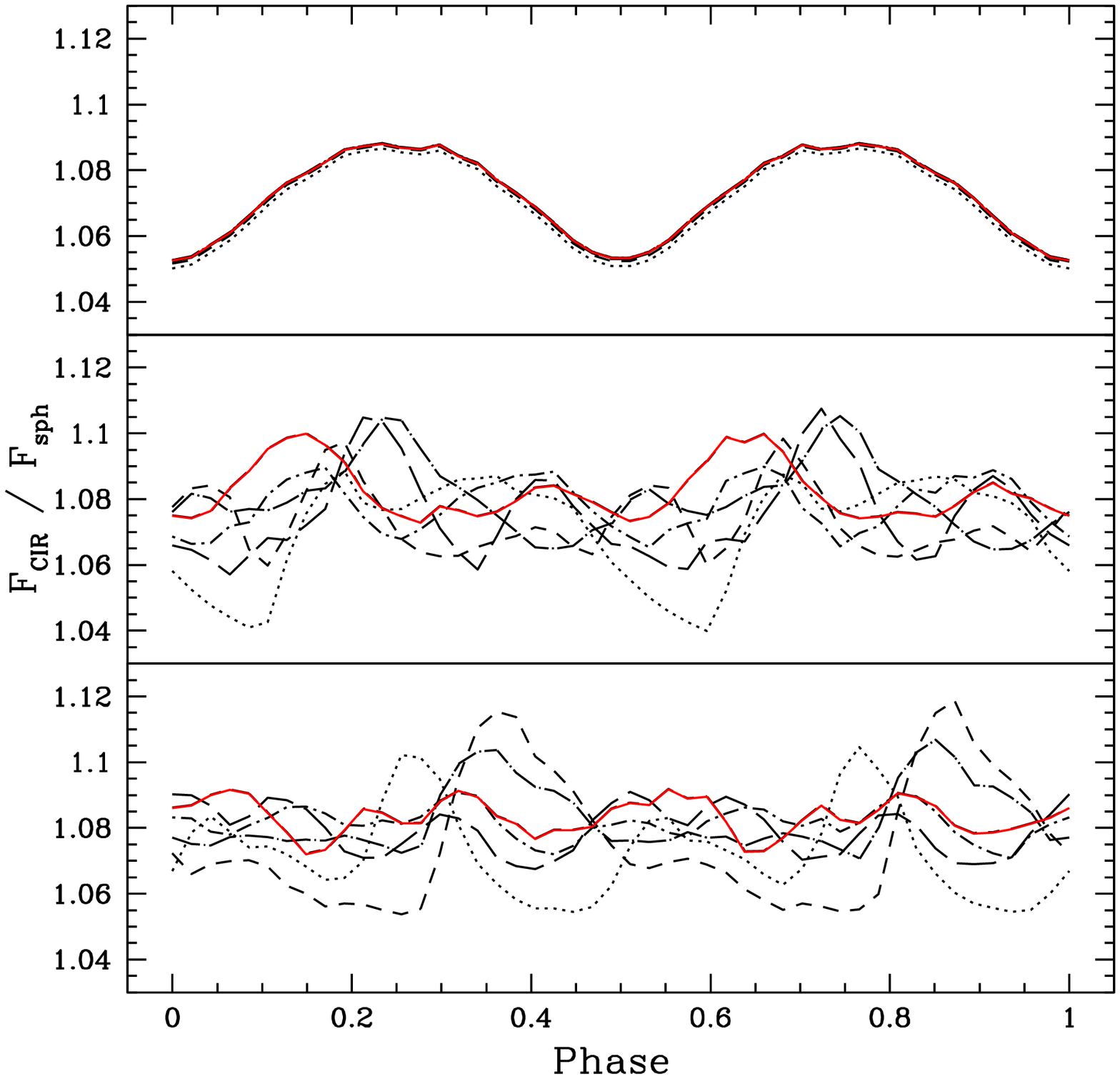}
\includegraphics[width=0.67\columnwidth]{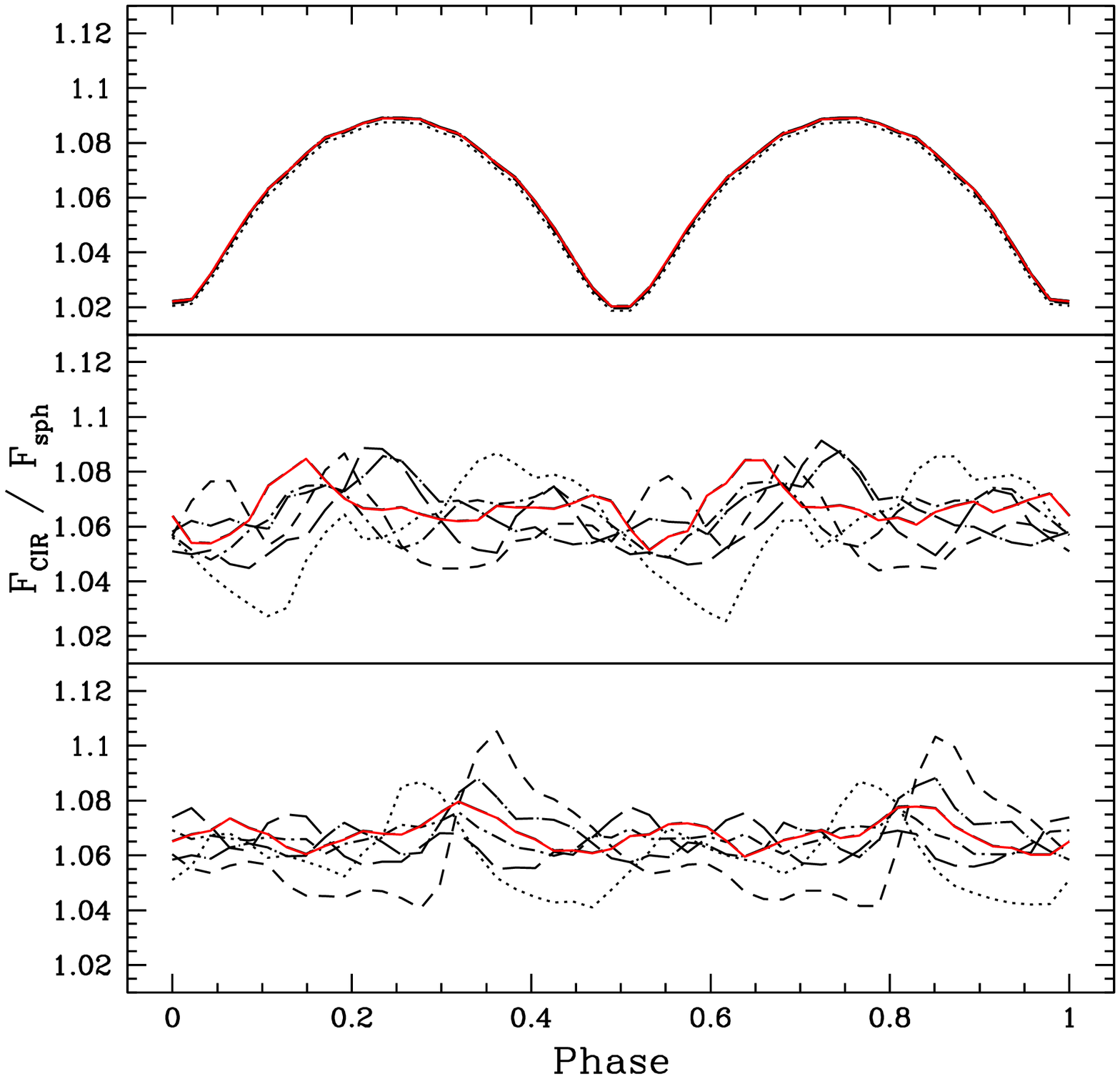}
 \caption{Same as Fig.~\ref{F1}, but with a different density contrast
 (see Tab.~\ref{tab1}).}
  \label{F2}
\end{figure*}

\section{Discussion}
\label{section3}

Using the model of the preceding section, we simulated
a large number of radio light curves.  The parameter space is 
extensive, including the half-opening angle $\beta _0$, the viewing inclination
$i _0$, the wind optical depth $\tau_0$, the co-latitude of the CIR
$\vartheta$,  the stellar rotation speed $v_{\rm rot}$ relative to the wind
speed, and the density contrast between the CIR and the wind $\eta$.  Note
that the latter is
defined as

\begin{equation}
\eta = \frac{n_{CIR}-n_{\rm sph}}{n_{\rm sph}},
\end{equation}

\noindent where $n_{CIR}$ is the number density inside the CIR structure, and
$n_{\rm sph}$ is the number density in the otherwise spherical wind.  It is
assumed that $\eta$ is a constant across the CIR, and with distance from the
star.  In addition to all these parameters, one further expects the detailed characteristics
of the radio light curve to be a function of wavelength.  Finally,
one could even allow for multiple
CIRs in the wind.

To keep the parameter space manageable, we have adopted a few
simplications.  First, we consider a wind with just one
equatorial CIR, hence $\vartheta=90^\circ$.  Second, noting that a pole-on view
produces no variation of flux with rotational phase,
we evaluate simulations for just three viewing
inclinations of $i_0=30^\circ$ (nearly pole-on), $60^\circ$
(mid-perspective), and $90^\circ$ (edge-on).  We consider
only two density contrasts of $\eta=3$ and 9.  We also
consider just two half-width opening angles of $\beta_0=
15^\circ$ and $25^\circ$.  

\begin{figure*}
\includegraphics[width=0.67\columnwidth]{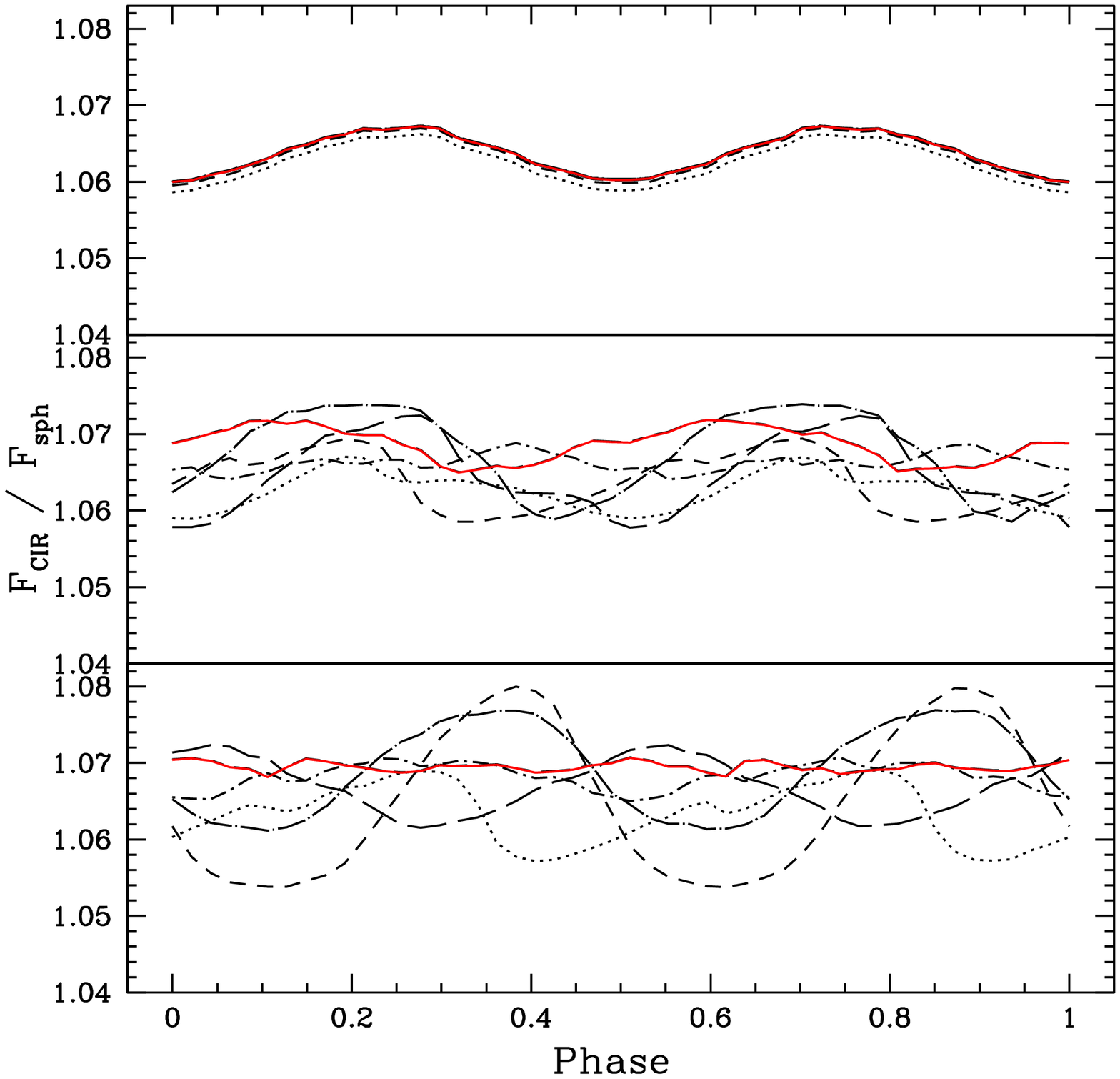}
\includegraphics[width=0.67\columnwidth]{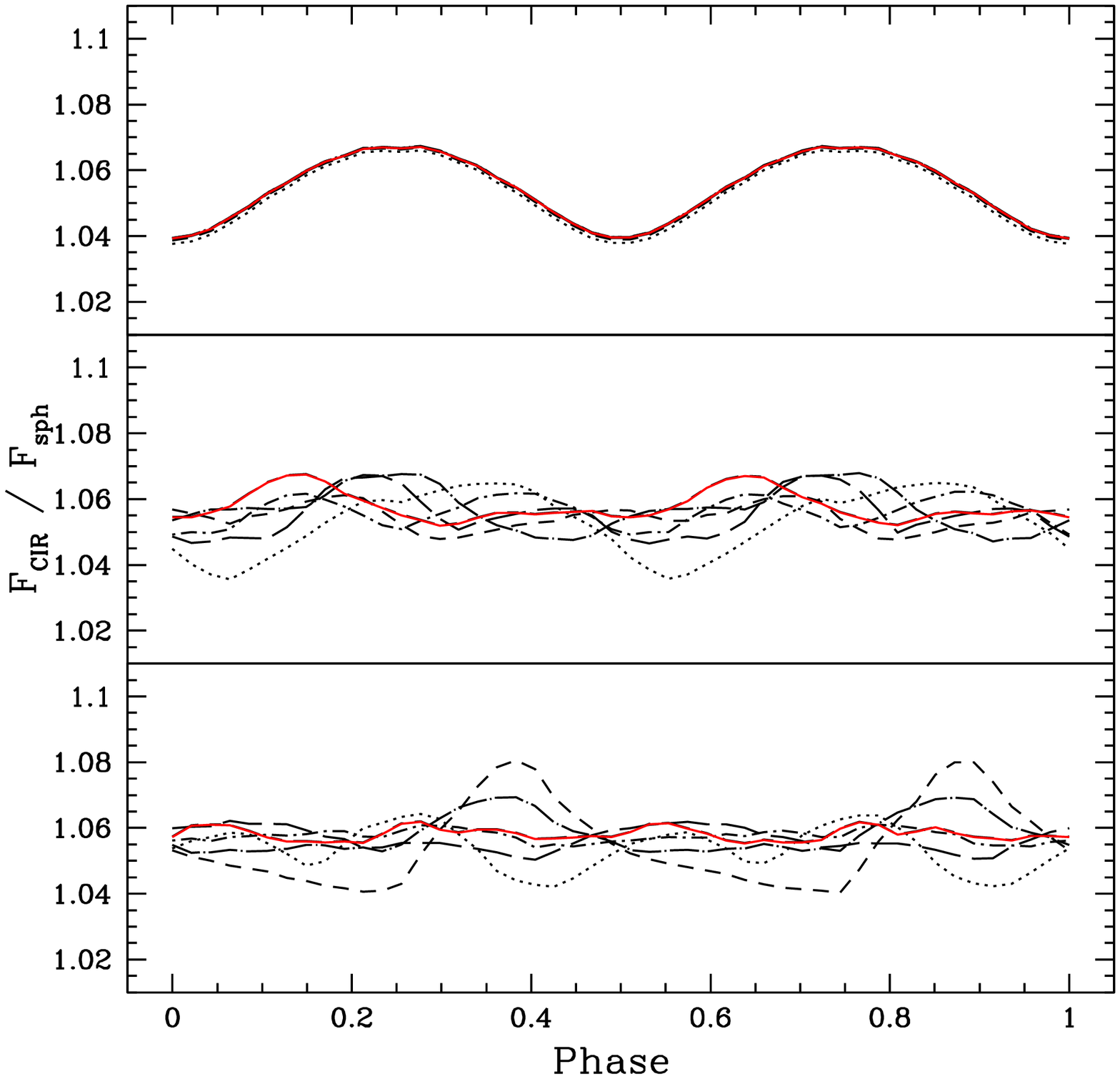}
\includegraphics[width=0.67\columnwidth]{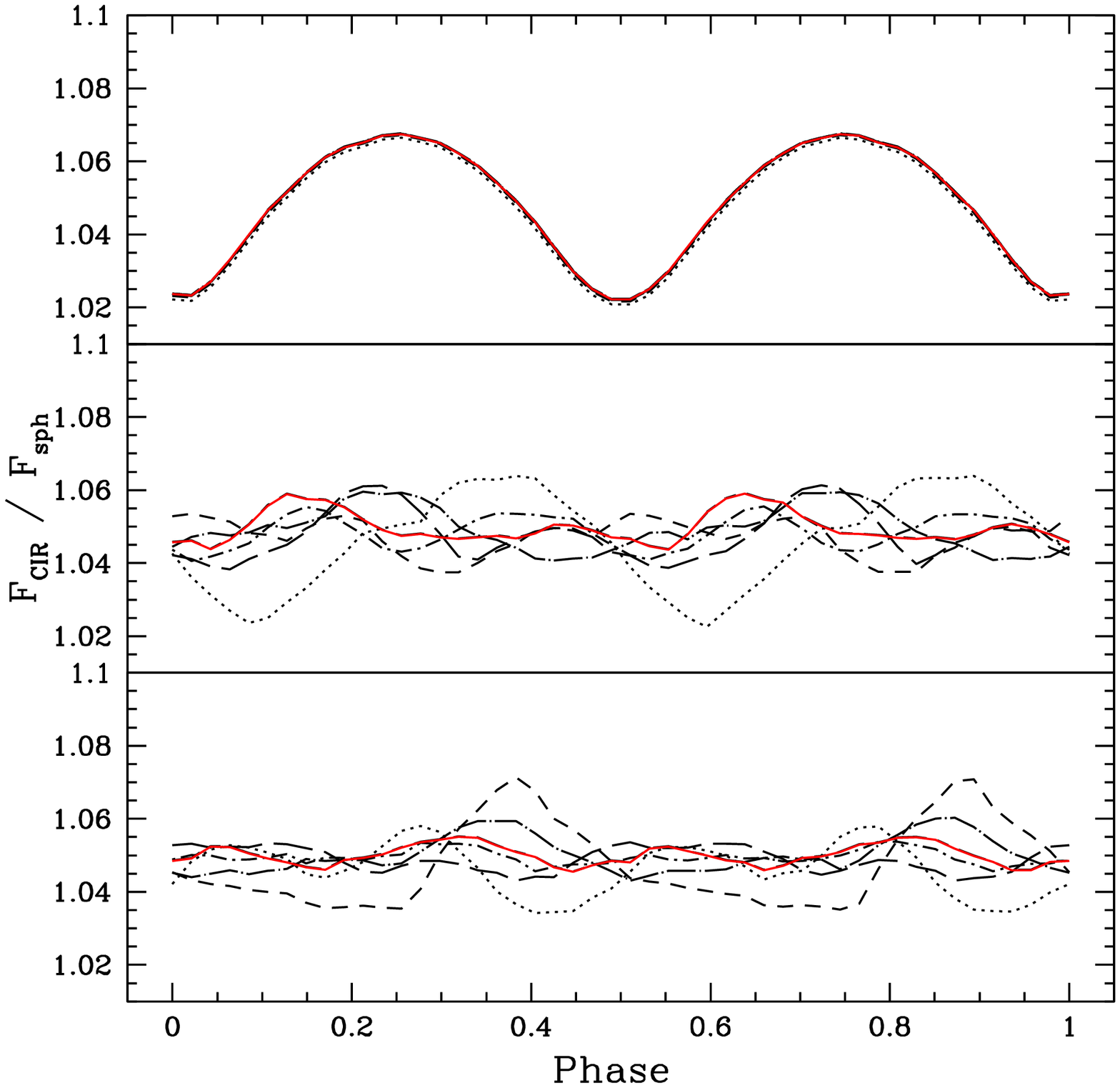}
 \caption{Same as Fig.~\ref{F1}, but now with a different opening angle
 (see Tab.~\ref{tab1}).}
  \label{F3}
\end{figure*}

\begin{figure*}
\includegraphics[width=0.67\columnwidth]{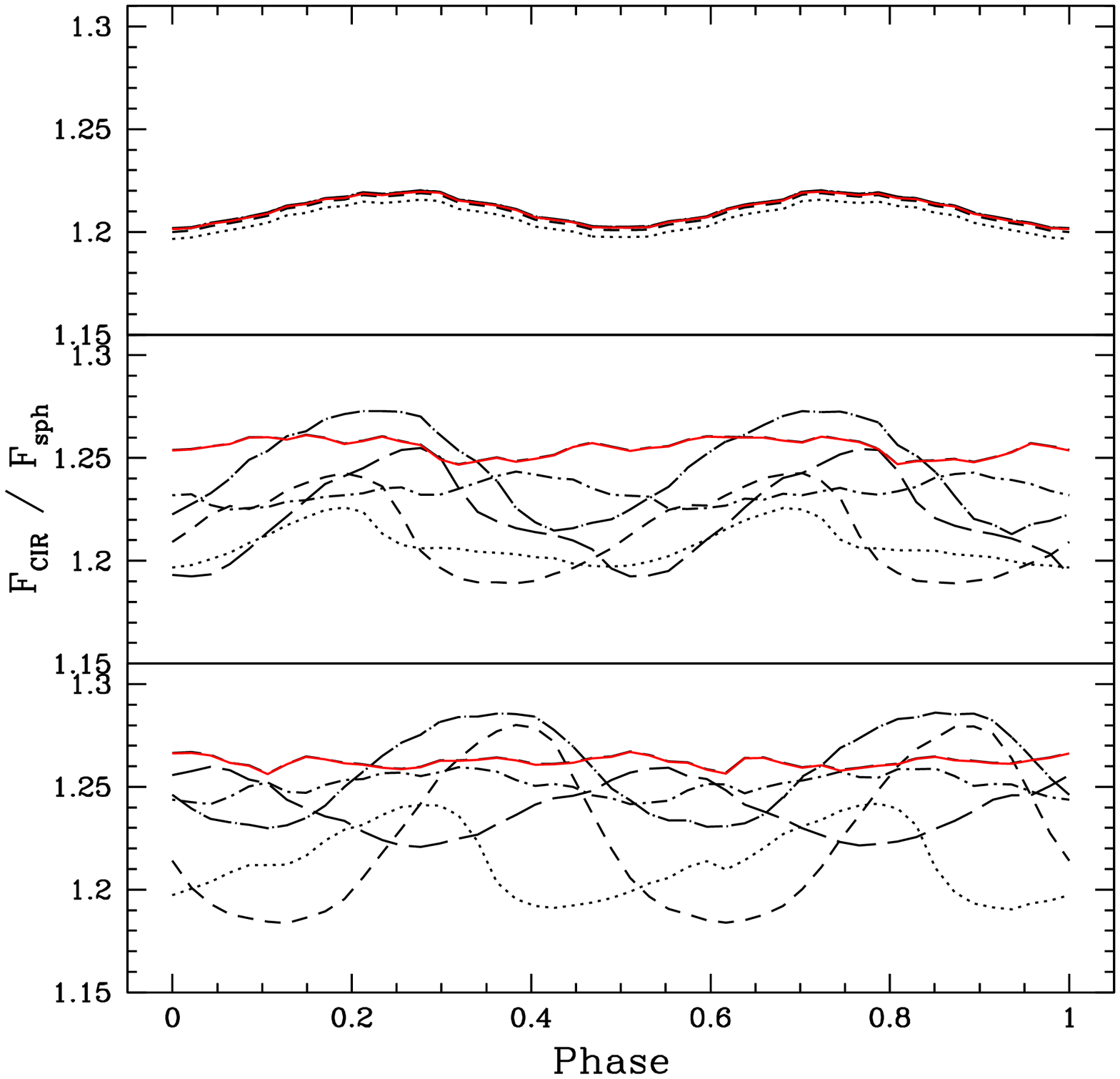}
\includegraphics[width=0.67\columnwidth]{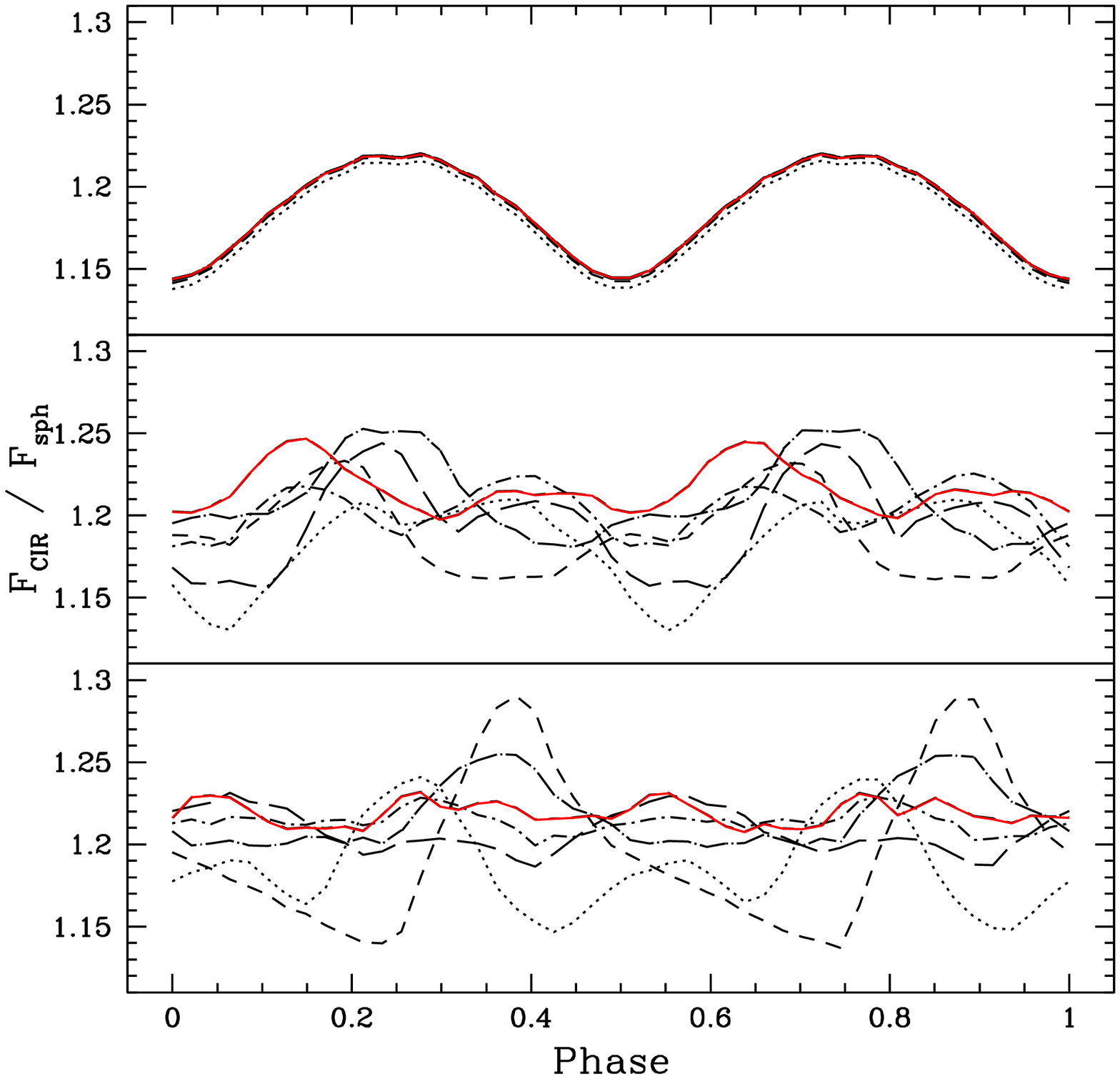}
\includegraphics[width=0.67\columnwidth]{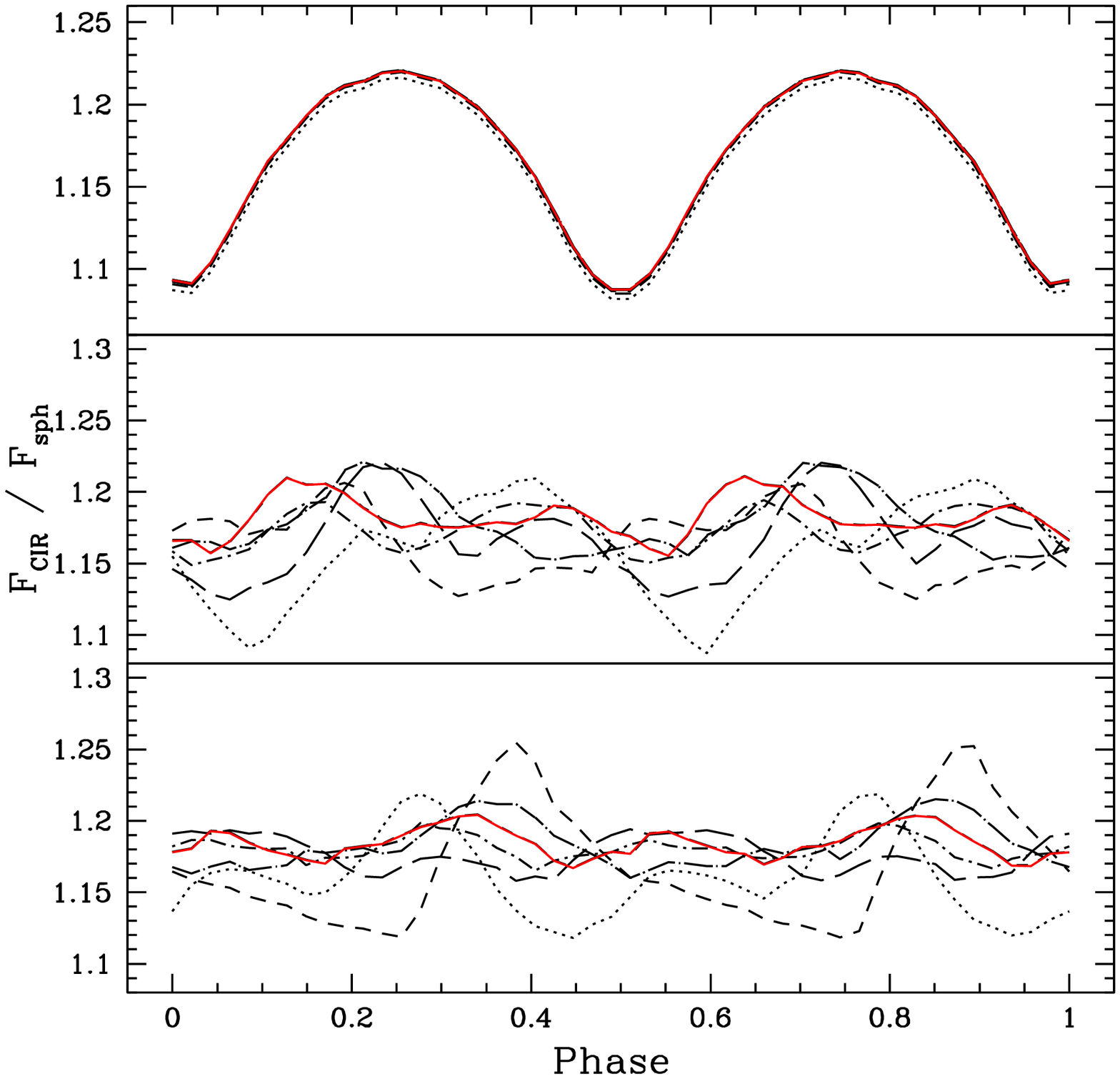}
 \caption{Same as Fig.~\ref{F3}, but now with a different density contrast
 (see Tab.~\ref{tab1}).}
  \label{F4}
\end{figure*}

Finally, we allow for three rotation speeds
of 0, 52, and 175 km s$^{-1}$.  A near-zero rotation speed represents
a strict conical CIR.  Consequently, the structure has no curvature,
so the perturbation to the wind is self-similar with wavelength.
The other two values represent low and medium rotations.
A fast rotation case is not included for reasons that will
soon be explained.  In all models the wind terminal
speed is fixed at $v_\infty=1750$ km s$^{-1}$, an intermediate
value among WR~stars, considering WN and WC subtypes.

\begin{figure}
\includegraphics[width=\columnwidth]{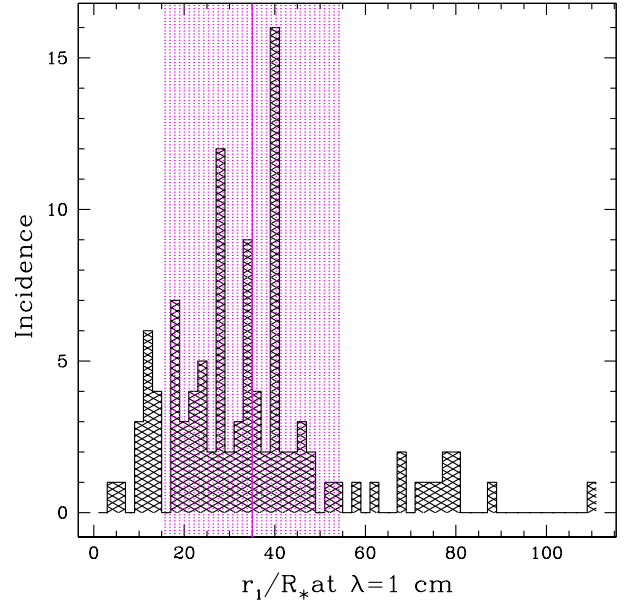}
 \caption{Shows the distribution of radio photospheres at $\lambda=1$ cm
 for single Galactic WN and WC stars.  The vertical magenta line is the average
 for the distribution.  The hashed magenta zone is $\pm 1$ standard deviation
 about the average for the distribution.  Refer to text for further details.}
  \label{F00}
\end{figure}

Figures~\ref{F1}-\ref{F4} display the model light curves and 
Table~\ref{tab1} details the various model parameters corresponding
to the figures.  In all cases, the spherical wind is assumed to have the same optical scale of $\tau_0 = 1.25\times 10^5$ at $\lambda=1$~cm.
This value was chosen as typical of WR stars.  The basis for this value derives from Figure~\ref{F00} which displays a histogram for $r_1/R_\ast$ for both WN and WC stars using stellar and wind parameters from \citet{2019A&A...625A..57H} and \citet{2019A&A...621A..92S} respectively. 
In calculating $r_1/R_\ast$ shown in this figure, representative values were adopted for mean molecular weights of WN
and WC stars.  

A typical volume filling factor of $f_V=0.1$ 
was  adopted.  Overall, an average value of
$\langle r_1/R_\ast \rangle = 35$ was determined.
While there is some tail in the histogram, there is a
fairly tight average value for the majority of stars.
Our selection of a single $\tau_0$ corresponds to this typical
value for $r_1/R_\ast$.

Turning back to the model light curves in Figures~\ref{F1}-\ref{F4},
each figure shows 3 sets of panels.  The left set is for $i=30^\circ$;
the center set is for $i=60^\circ$; and the right set is for
$i=90^\circ$.
For a given inclination, each of the 3 panels display radio light curves
for different rotational phases.  The top panel is for near-zero rotation
speed; middle is for the low speed case; bottom is for the modest
speed case.  Table~\ref{tab1} identifies the moderl parameters for
each panel, by identifying the figure, the set of panels (left, center, or
right), and the specific panel (top, middle or ``mid'', and bottom
or ``bot'').  All light curves are plotted as
fluxes normalized to what a strictly spherical wind would 
produce at the same wavelength.  Each panel has multiple
light curves for different wavelengths.  The wavelengths
range from 1~mm (dotted line) to 31.6~cm (red line)
in logarithmic intervals of 0.5~dex.


\begin{figure*}
\includegraphics[width=1.4\columnwidth]{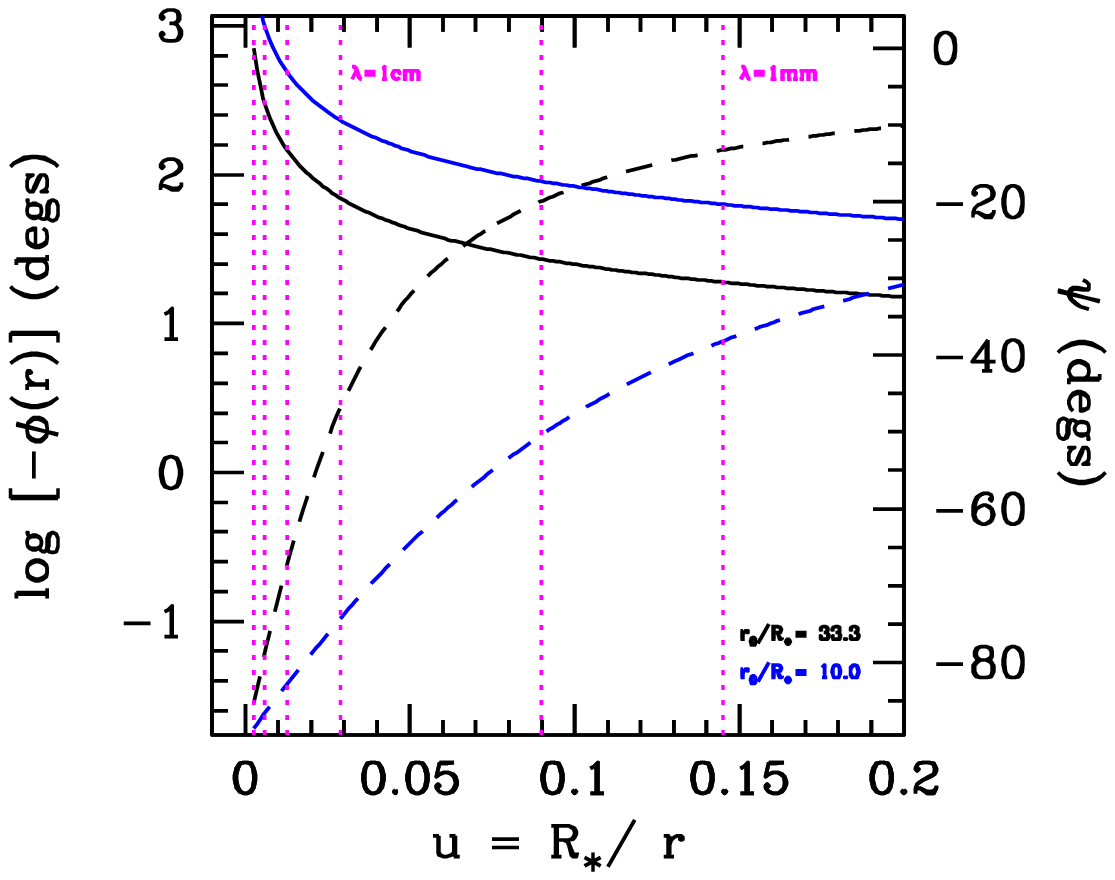}
 \caption{The left axis shows the wrapping of the CIR, $\phi$, in degrees
 in the log shown as the solid curve.  The black and blue curves are for winding
 radii, $r_0/R_\ast$, as indicated; these correspond to the slower and faster
 rotation speeds associated with the light curves of Figs.~\ref{F1}-\ref{F4}.
 Note that 1 full wrapping of the CIR about the star is $360^\circ$ or
 about 2.56 in the log.  
 The right axis is for the orientation of the unit tangent vector
 to the CIR, with $\psi=0^\circ$ being radial and $\psi= -90^\circ$ being
 backward azimuthal.  The solutions for $\psi$ are displayed as dashed lines.
 The vertical magenta lines are for $r_1/R_\ast$ at the 6 wavelengths used
 for the light curve simulations, with the 1~mm and 1~cm lines labeled.}
  \label{F13}
\end{figure*}

The results of the simulations can be summarized in
four main points:

\begin{enumerate}

\item In all cases a CIR produces a flux offset.  This is the same effect that
stochastic clumping would have on the wind.  In effect, a CIR can be considered as a clump with an organized
geometry that produces systematic effects with rotation.  The extent of the
excess relative to a spherical wind scales with the gross properties of the CIR,
such as density contrast $\eta$ and opening angle.  For the parameters used in this study,
excesses range from a few percent up to a few tens of percent.  However, in the radio
band, the presence of a CIR is betrayed through cyclic radio variability,
with detectability set by the peak-to-trough variations of the excess.  Those
variations tend to be on the order of 10\% or less, again for the adopted 
parameters.

\item The top panels of each figure is for a conical CIR.  The assumption here
is that rotation is so slow that the CIR has essentially no curvature.
Two points are notable.  First, we expect two peaks per
rotation corresponding to when the CIR is in the plane of the sky.  The simulations
adopt zero phase as being when the footpoint of the CIR
is on the near-side of the star.
Then phase of 0.5 has the footpoint rear of the star; 0.25 has the footpoint
in the plane of sky; and 0.75 is also in the sky plane but opposite of 0.25.
The sense of rotation is counterclockwise as seen from above, hence if viewed
edge-on, phase of 0.25 is rightside in the sky, and 0.75 is leftside, with the 
rotation axis lying in the plane of the sky and directed up to define right and left.
Recall that the radio photosphere grows in extent with increasing wavelength.
The absence of curvature for a conical CIR means that at every wavelength,
the light curve samples the exact same relative geometry (i.e., the CIR is
azimuthally aligned
with its footpoint at all radii).

\item The middle and bottom panels allow for curvature of the CIR.  With increasing
radius, the CIR is becoming more ``wound up''.  At short wavelengths the radio 
photosphere forms where the CIR is less wound up; at long wavelengths, the photosphere
is sampling a CIR that has a greater degree of spiral morphology, which depends
on the ratio $v_{\rm rot}/v_\infty$.  In general, this means that the emergence
of the CIR at the wavelength-dependent photosphere becomes increasing lagged
in azimuth as compared to the CIR footpoint.  For example, when the footpoint is
at a phase of 0.25, the CIR is backwinding and could emerge at the radio photosphere
more in front of the star.  

\item The asymptotic effect of a highly wound CIR is no longer to produce a variation
with rotational phase.  There is still an excess of flux above the spherical value, yet 
the light curve develops a variety of small peaks and troughs.  Variability is
suppressed because many wrappings of the CIR means that the projected radio photosphere,
if it could be resolved, looks mostly the same at every wavelength.  The many
wrappings approximate a structure that becomes nearly axisymmetric.  This can be seen
in long-wavelength light curves (red) for the bottom panels.  And it is why 
fast rotation speed cases were not included in the study, since these would only give low-variability behavior.

\end{enumerate}

To illustrate how different wavelengths probe different
geometries of the wind, Figure~\ref{F13} displays the radii
of the radio photospheres for the six wavelengths used
with the simulated light curves in relation to how wrapped
up the CIR has become.  The horizontal axis is inverse
radius, $u$, in the wind.  The vertical magenta lines
are the $u_1 = R_\ast/r_1(\lambda)$ values for the six
wavelengths, with 1~mm and 1~cm cases labeled.  The leftside
axis is for the azimuth of the center of the CIR as a function
of radius from equation~(\ref{eq:kinematics}).  However, it is plotted
here in the log of $-\phi$, since by convention the CIR is backwinding
relative to the stellar rotation.  (Note that $-\phi = 360^\circ$
is 2.56 in the log.)  The rightside axis is for the orientation,
$\psi$ of the local tangent to the CIR, with

\begin{equation}
    \tan \psi = \frac{v_\phi(r) - \Omega\,r}{v_{\rm r}}.
\end{equation}

\noindent If the CIR were radial, $\psi=0^\circ$; if it
were purely azimuthal, $\psi= -90^\circ$, with the negative
appearing because the CIR is backwinding.

Figure~\ref{F13} shows two sets of curves.  Black and blue
colors are for values of the winding radius, $r_0$, as
indicated in the figure.  Solid is for $\phi$ and dashed
is for $\psi$.  The figure makes clear that owing to the
increasing opacity with wavelength, radio data probe different
degrees of curvature of the CIR, possibly even multiple 
wrappings, with $\lambda$.  

\section{Conclusion}
\label{section4}

There is unambiguous evidence today that radiatively driven winds
are far more complex than the homogeneous, spherically symmetric
flows originally envisioned \citep[e.g.][]{1975ApJ...195..157C}.
Instead, they have been shown to contain optically thick structures
which may be quite small (micro-structures) or very large
(macro-structures). Unraveling the details of these flows is a
prerequisite to translating observational diagnostics into reliable
physical quantities such as mass-loss rates. To progress, a firm
grasp of the underlying physical mechanisms that determine the wind
structures is needed. Specifically, the passage of CIR spiral arms
across the line of sight to the stellar disk accounts for the wind
line UV variability, and in that context they undoubtedly strongly
affect the observational diagnostics used to determine the true
mass-loss rates.  The CIR density enhancements also provide a
potentially powerful, but untested, means for producing radio
variability.  

We have demonstrated here that {\it temporal} radio continuum
(multi-frequency) datasets can potentially provide a powerful new
key to developing a coherent picture of wind flows, their large-scale
structure and how they fit together.  It is important to note
that several simplifications have been invoked in order to facillitate
a broad parameter study in terms of CIR geometry, density compression,
and viewing inclination along with creating simulated multi-wavelength
light curves.  We recognize that hydrodynamic models for CIRs show
not only compressions but also rarefactions.  We anticipate a
follow-up study to explore the impact of rarefactions for the light
curves.  A sector of depressed density acts in opposition to a
sector where density is enhanced.  The latter extends the radio
photosphere; the former contracts it.  One may expect that a
rarefaction will lower the overall radio excess of a wind with a
CIR as compared to a spherical wind.  Correspondingly, variability
is driven by how the projected radio photosphere changes with
rotational phase.  With both extension and contraction available,
inclusion of rarefaction may increase the relative variability,
although this is likely sensitive to viewing inclination.

What has been demonstrated is that new perspectives on large-scale
wind structure can be provided by monitoring the radio continuum
emission of WR and luminous O stars simultaneously at multi-radio
bands (e.g., 1~cm, 6~cm, and 21~cm) over CIR (essentially stellar
rotation) timescales.  The variability from a CIR derives from a
phase dependence of the projected and non-centrosymmetric photosphere
with rotation. For an equatorial CIR that is quite over-dense
compared to its surroundings, the overall variation in flux for the
unresolved source could achieve variations of 10--20\%.  From an
observational perspective, curvature due to a spiral-CIR results
in the development of a potentially exploitable phase lag, provided
the bands have sufficient dynamic range in wavelength and are
monitored contemporaneously.  Another prediction of the model is
that the amplitude decreases as the photosphere grows with wavelength,
because more winding up of the spiral leads to less relative
variability.  So at short wavelength, there are two peaks because
the CIR is more nearly like a cone and results in maximum flux
excess when the cone is on either side of the star in the plane of
sky.  By contrast, at truly long wavelengths, which diagnose a CIR
that is very wrapped up, there is indeed a flux enhancement but no
variability.

Ultimately, powerful radio astronomy facilities such as the Square
Kilometre Array (SKA) will open up new avenues such time-domain
surveys of massive star winds.  The interpretation of these very
rich datasets will require an understanding of the contribution of
large-scale wind structures to the thermal and non-thermal emission
in the radio.

\section*{Acknowledgements}
The authors express appreciation to the anonymous referee for raising
several points that have improved this paper.  RI acknowledges
support by the National Science Foundation under Grant No. AST-1747658.
NSL wishes to thank the National Sciences and Engineering Council
of Canada (NSERC) for financial support.

\section*{Data Availability Statement}

The data underlying this article will be shared on reasonable request
to the corresponding author.




\bibliographystyle{mnras}
\bibliography{st-louis} 





\appendix



\bsp	
\label{lastpage}
\end{document}